\documentclass[aps,eqsecnum]{revtex4}
\usepackage{graphicx}
\begin{document}
\def\ii{\'{\i}}
\title{Hybrid stars in the quark-meson coupling model with superconducting 
quark matter}
\author{P.K. Panda}
\affiliation{Depto de F\ii sica - CFM - Universidade Federal de Santa
Catarina  Florian\'opolis - SC - CP. 476 - CEP 88.040 - 900 - Brazil}
\author{D.P. Menezes}
\affiliation{Depto de F\ii sica - CFM - Universidade Federal de Santa
Catarina  Florian\'opolis - SC - CP. 476 - CEP 88.040 - 900 - Brazil}
\author{C. Provid\^encia}
\affiliation{Centro de F\ii sica Te\'orica - Dep. de F\ii sica - 
Universidade de Coimbra - P-3004 - 516 Coimbra - Portugal}
\begin{abstract}
The phase transition to a deconfined phase is studied and the consequences in 
the formation of neutron stars are investigated. 
We use the quark-meson-coupling model for the hadron matter and the MIT Bag 
model for the quark matter in order to build the appropriate equations of state
for the hybrid stars. The properties of the stars are then calculated.
The differences between unpaired and color-flavor locked quark matter are 
discussed.
\end{abstract}
\date{\today}
\maketitle
\vspace{0.2cm}
PACS number(s): {95.30.Tg, 21.65.+f, 12.39.Ba, 24.85+p, 21.30.-x}
\vspace{0.2cm}
\section{Introduction}
It is widely believed that hadronic matter undergoes a phase
transition to quark matter at high densities and/or high temperatures. The 
high temperature limit is important in heavy ion collisions and/or
cosmology, whereas the high baryon density behavior is important for
the study of neutron, hybrid and quark stars.

After a gravitational colapse of a massive star takes place, a neutron star
with practically zero temperature can be born few seconds after
deleptonization.
The correct calculations of the star properties depend on the appropriate 
equations of state (EOS) that describe its crust and interior. The crust of 
the 
neutron star, where density is low is believed to be adequately described by
hadronic matter. Its interior, however, where density is of the order of  
$5 \sim 10 $
times nuclear saturation density remains to be properly understood. Whether 
the central part of the star is composed of quark matter only, of mixed matter
or of paired quark matter is one of the subjects of the present work. 

In the present paper we are interested in building the EOS for mixed 
matter of quark and hadron phases. We employ the quark-meson coupling 
model (QMC)~\cite{guichon,ST} 
including hyperons in order to describe the hadron phase. In the QMC model, 
baryons are described as a system of non-overlapping MIT bags which interact
through the effective scalar and vector mean fields, very much in the same
way as in the Walecka model (QHD) \cite{qhd}. Many applications and extensions 
of the model have been made in the last years 
\cite{tsht,pal,recent,temp,phase,jin96,mj98}.

While the QMC model shares many similarities 
with QHD-type models, it however offers new opportunities for studying nuclear 
matter properties. One of the most attractive aspects of the model is that 
different phases of hadronic matter, from very low to very 
high baryon densities 
and temperatures, can be described within the same underlying model,
namely the MIT bag model.
In the QMC, matter at low densities and temperatures is a system of nucleons 
interacting through meson fields, with quarks and gluons confined within 
MIT bag. For matter at 
very high density and/or temperature, one expects that baryons and 
mesons dissolve and the entire system of quarks and gluons becomes 
confined within a single, big MIT bag. Another important aspect of the 
QMC is that the internal structure of the nucleon is introduced explicitly. 
It is found that the EOS for infinite 
nuclear matter at zero temperature derived from the QMC model is much softer 
than the one obtained in  the Walecka model~\cite{qhd}. Also, the QMC model 
nucleon 
effective mass lies in the range $ 0.7$ to $0.8$ of the free nucleon mass, 
which agrees with results derived from non-relativistic analysis of scattering 
of neutrons from lead 
nuclei~\cite{mahaux87} and is  larger in comparison with Walecka model 
effective 
mass. Consequently, we expect that contrary to the NL3 and TM1 
parameterizations of the non linear Walecka model (NLWM) with hyperons, the 
mass of the nucleon does not become zero at densities $\rho<10\rho_0$, 
$\rho_0$ being the density of nuclear matter at saturation.
 At finite temperature, there arises yet another difference between
predictions of QMC and QHD, namely the behavior of the effective nucleon mass 
with the temperature at fixed baryon density. While in QHD-type models the 
nucleon mass always decreases with temperature, in the QMC it increases. The 
difference arises because of the explicit treatment of the internal structure 
of the nucleon in the QMC. When the bag is heated up,  
quark-antiquark pairs are excited in the interior of the bag, increasing the 
internal energy of the bag \cite{phase}.  
In what follows we consider the QMC with a constant bag constant, e. g. 
not dependent on the scalar meson field as in the modified quark meson 
coupling models (MQMC) \cite{jin96,mj98}.Contrary to the QMC
result, the bag radius increases with density for all MQMC models with a 
meson dependent bag constant \cite{mj98}. For densities not much larger than 
nuclear matter saturation density $\rho_0$ the bags start to overlap which 
implies a breakdown of the model.  In the present work 
we are interested in describing properties of nuclear matter for
densities which go  beyond the saturation density and therefore have chosen to
consider the QMC model.

For the quark phase we have chosen to use both unpaired quark matter (UQM) 
described by the MIT bag model \cite{bag,freedman,farhi,kapusta} 
and paired quarks described by the color-flavor locked (CFL) phase. 
Recently many authors \cite{shh,ar,bo,srp,arrw,kw} have discussed the 
possibility that the quark matter is in a color-superconducting phase, in 
which quarks near the Fermi surface are paired, forming Cooper pairs which 
condense and break the color gauge symmetry \cite{mga}. At sufficiently high 
density the favored phase is called CFL, in which quarks of all three colors 
and all three flavors are allowed to pair. 

Once the mixed EOS is built by
enforcing appropriate Gibbs criteria and chemical equilibrium conditions, the
properties of the stars are calculated and discussed. We have restricted 
ourselves to the 
investigation of neutron stars ($T=0$) and hence, neutrino trapping is not 
an expected mechanism because 
it is important only during the first seconds after gravitational
collapse of the core of a massive star \cite{bl86,kj95}.

In recent works \cite{recente1,recente2}, two of us have studied the 
properties of mixed stars whose equations of state were built with
the non-linear Walecka model (NLWM) for the hadron matter \cite{qhd} and 
the MIT Bag \cite{bag,freedman,farhi,kapusta} and the Nambu-Jona-Lasinio 
\cite{njl} models for the unpaired quark matter. In the first work 
\cite{recente1} the effects of temperature 
were investigated and in the second one the consequences of imposing neutrino 
trapping at fixed entropies were studied. In both works, only in very special 
cases, the interior of the stars were made of quarks only. 
In general, a mixed phase of hadrons and quarks was favored.

In what follows we compare the properties of neutron stars obtained within 
the QMC model with hyperons and with both 
quark models discussed above, namely, the onset of hyperons, mixed and quark 
phases, strangeness content, the maximum allowed masses and core composition.

The present paper is organized as follows: in section II the QMC model
with hyperons is reviewed. In sections III and IV the unpaired quark matter 
and the CFL phase are described. In section V the mixed 
phase is implemented and the results are shown and discussed in section VI. 
Finally, in the last section, the conclusions are drawn. 

\section{The quark-meson coupling model for hadronic matter}

In the QMC model, the nucleon in nuclear medium is assumed to be a
static spherical MIT bag in which quarks interact with the scalar and
vector fields, $\sigma$, $\omega$ and $\rho$ and these
fields are treated as classical fields in the mean field
approximation.  The quark field, $\psi_q(x)$, inside the bag then 
satisfies the equation of motion: 
\begin{equation}
\left[i\,\rlap{/}\partial-(m_q^0-g_\sigma^q\, \sigma)
-g_\omega^q\, \omega\,\gamma^0 + \frac{1}{2} g^q_\rho \tau_z \rho_{03}\right]
\,\psi_q(x)=0\ , \quad  q=u,d,s,
\label{eq-motion}
\end{equation}
where $m_q^0$ is the current quark mass, and $g_\sigma^q$,
$g_\omega^q$ and $g_\rho^q$ and denote the quark-meson coupling constants. 
The normalized ground state for a quark in the bag is given
by \cite{guichon,ST}
\begin{equation}
\psi_q({\bf r}, t) = {\cal N}_q \exp \left(-i\epsilon_q t/R_B \right)
\left( \matrix{ j_0\left(x_q r/R_B\right)\cr
i\beta_q \vec{\sigma} \cdot \hat r j_1\left(x_q r/R_B\right) }
\right) \frac{\chi_q}{\sqrt{4\pi}} ~,
\end{equation}
where 
\begin{equation}
\epsilon_q=\Omega_q +R_B\left(g_\omega^q\, \omega+
\frac{1}{2} g^q_\rho \tau_z \rho_{03} \right)  ~; ~~~
\beta_q=\sqrt{{\Omega_q-R_B\, m_q^*\over \Omega_q\, +R_B\, m_q^*}}\ ,
\end{equation}
with the normalization factor given by
\begin{equation}
{\cal N}_q^{-2} = 2R_B^3 j_0^2(x_q)\left[\Omega_q(\Omega_q-1)
+ R_B m_q^*/2 \right] \Big/ x_q^2 ~,
\end{equation}
where $\Omega_q\equiv \sqrt{x_q^2+(R_B\, m_q^*)^2}$, 
$m_q^*=m_q^0-g_\sigma^q\, \sigma$, $R_B$ is the bag radius of the baryon, 
and  $\chi_q$ is the quark spinor. The quantities $\psi_q,\, \epsilon_q,\, 
\beta_q,\, {\cal N}_q,\, \Omega_q,\, m^*_q$ all depend on the baryon 
considered. The bag eigenvalue, $x_q$, is determined by the 
boundary condition at the bag surface
\begin{equation}
j_0(x_q)=\beta_q\, j_1(x_q)\ .
\label{bun-con}
\end{equation}
The energy of a static bag describing baryon $B$ consisting of three ground state quarks 
can be expressed as
\begin{equation}
E^{\rm bag}_B=\sum_q n_q \, {\Omega_q\over R_B}-{Z_B\over R_B}
+{4\over 3}\,  \pi \, R_B^3\,  B_B\ ,
\label{ebag}
\end{equation}
where $Z_B$ is a parameter which accounts for zero-point motion
and $B_B$ is the bag constant.
The set of parameters used in the present work is given 
in table \ref{tab1}. The effective mass of a nucleon bag at rest
is taken to be
\begin{equation}
M_B^*=E_B^{\rm bag}.
\label{eff-mn}
\end{equation}
The equilibrium condition for the bag is obtained by 
minimizing the effective mass, $M_B^*$ with respect to the bag radius
\begin{equation}
{d\, M_B^*\over d\, R_B^*} = 0\ .
\label{balance}
\end{equation}
For the QMC model, the equations of motion for the meson fields in
uniform static matter are given by
\begin{equation}
m_\sigma^2\sigma = \sum_B g_{\sigma B} C_B(\sigma) \frac{2J_B + 1}{2\pi^2}
\int_0^{k_B} \frac{M_B^*(\sigma)}
{\left[k^2 + M_B^{* 2}(\sigma)\right]^{1/2}} \: k^2 \ dk ~,
\label{field1}
\end{equation}
\begin{equation}
m_\omega^2\omega_0 = \sum_B g_{\omega B} \left(2J_B + 1\right)
k_B^3 \big/ (6\pi^2) ~,
\label{field2}
\end{equation}
\begin{equation}
m_\rho^2\rho_{03} = \sum_B g_{\rho B} I_{3B} \left(2J_B + 1\right)
k_B^3 \big/ (6\pi^2) ~.
\label{field3}
\end{equation}
In the above equations $J_B$, $I_{3B}$ and $k_B$ are respectively the spin, 
isospin projection and the Fermi momentum of the baryon species $B$.
The hyperon couplings are not
relevant to the ground state properties of nuclear matter,but information
about them can be available from the levels in $\Lambda$ hypernuclei 
\cite{chrien}.  
$$g_{\sigma B}=x_{\sigma B}~ g_{\sigma N},~~g_{\omega B}=x_{\omega B}~ 
g_{\omega N}, ~~g_{\rho B}=x_{\rho B}~ g_{\rho N}$$
and $x_{\sigma B}$, $x_{\omega B}$ and $x_{\rho B}$ are equal to $1$ for the 
nucleons and acquire different values in different parameterizations for the 
other baryons. Note that the $s$-quark is unaffected by the sigma and omega
mesons i.e. $g_\sigma^s=g_\omega^s=0\ .$

In Eq. (\ref{field1}) we have
\begin{equation}
g_{\sigma B}C_B(\sigma) = - \frac{\partial M_B^*(\sigma)}{\partial \sigma} 
= - \frac{\partial E^{\rm bag}_B}{\partial \sigma} 
= \sum_{q=u,d} n_q g^q_\sigma S_B(\sigma)
\end{equation}
where
\begin{equation}
S_B(\sigma) = \int_{bag} d{\bf r} \ {\overline \psi}_q \psi_q
= \frac{\Omega_q/2 + R_Bm^*_q(\Omega_q - 1)}
{\Omega_q(\Omega_q -1) + R_Bm_q^*/2} ~; ~~~~ q \equiv (u,d) ~.
\end{equation}

The total energy density and the pressure
including the leptons can be obtained from the grand canonical
potential and they read
\begin{eqnarray}
\varepsilon &=& \frac{1}{2}m_\sigma^2 \sigma^2
+ \frac{1}{2}m_\omega^2 \omega^2_0
+ \frac{1}{2} m_\rho^2 \rho^2_{03} \nonumber\\
&+& \sum_B \frac{2J_B +1}{2\pi^2} \int_0^{k_B}k^2 dk
\left[k^2 + M_B^{* 2}(\sigma)\right]^{1/2} 
+ \sum_l \frac{1}{\pi^2} \int_0^{k_l} k^2  dk\left[k^2 + m_l^2\right]^{1/2}~,
\end{eqnarray}
\begin{eqnarray}
P &=& - \frac{1}{2}m_\sigma^2 \sigma^2
+ \frac{1}{2}m_\omega^2 \omega^2_0
+ \frac{1}{2} m_\rho^2 \rho^2_{03} \nonumber\\
&+& \frac{1}{3} \sum_B \frac{2J_B +1}{2\pi^2} \int_0^{k_B}
\frac{k^4 \ dk}{\left[k^2 + M_B^{* 2}(\sigma)\right]^{1/2}}
+ \frac{1}{3} \sum_l \frac{1}{\pi^2} \int_0^{k_l} \frac{k^4  dk}
{\left[k^2 + m_l^2\right]^{1/2}} ~.
\end{eqnarray}

The lepton Fermi momenta are the positive real solutions of
$(k_e^2 + m_e^2)^{1/2} =  \mu_e$ and
$(k_\mu^2 + m_\mu^2)^{1/2} = \mu_\mu = \mu_e$. The equilibrium composition
of the star is obtained by solving the set of Eqs. (\ref{field1})-
(\ref{field3}) in conjunction
with the charge neutrality condition  (\ref{neutral})  at a given total baryonic density
$\rho = \sum_B (2J_B + 1) k_B^3/(6\pi^2)$; the baryon effective masses are
obtained self-consistently in the bag model. 

For stars in which the strongly interacting particles are baryons, the
composition is determined by the requirements of charge neutrality
and $\beta$-equilibrium conditions under the weak processes
$B_1 \to B_2 + l + {\overline \nu}_l$ and $B_2 + l \to B_1 + \nu_l$.
After deleptonization, the charge neutrality condition yields
\begin{equation}
q_{\rm tot} = \sum_B q_B (2J_B + 1) k_B^3 \big/ (6\pi^2)
+ \sum_{l=e,\mu} q_l k_l^3 \big/ (3\pi^2)  = 0 ~,
\label{neutral}
\end{equation}
where $q_B$ corresponds to the electric charge of baryon species $B$
and $q_l$ corresponds to the electric charge of lepton species $l$. Since
the time scale of a star is effectively infinite compared to the weak
interaction time scale, weak interaction violates strangeness conservation.
The strangeness quantum number is therefore not conserved
in a star and the net strangeness is determined by the condition of
$\beta$-equilibrium which for baryon $B$ is then given by
$\mu_B = b_B\mu_n - q_B\mu_e$, where $\mu_B$ is the chemical potential
of baryon $B$ and $b_B$ its baryon number. Thus the chemical potential of any
baryon can be obtained from the two independent chemical potentials $\mu_n$
and $\mu_e$ of neutron and electron respectively.
 
We start by fixing the free-space bag properties for the QMC model. 
For the bag radius $R_N=0.6$, we first fixed the two unknowns $Z_N$ and $B_N$
for nucleons. These are obtained by fitting the neuclon mass $M=939$ MeV and
enforcing the stability condition for the bag at free space. The values
obtained are $Z_N=3.98699$ and $B_N^{1/4}=211.303$ MeV  for $m_u=m_d=0$ MeV and
$Z_N=4.00506$ and $B_N^{1/4}=210.854$ MeV for $m_u=m_d=5.5$ MeV. We then fixed
these bag values, $B_B$, for all baryons and the parameter $Z_B$ and $R_B$
of the other baryons are obtained
by reproducing their physical masses in free space and again enforcing
the stability condition for their bags. The values obtained for 
$Z_B$ and $R_B$ are displayed in table I for $m_u=m_d=0$  MeV 
and for $m_u=m_d=5.5$ MeV.
Note that for a fixed bag 
value, the equilibrium condition in free space results in an increase of the 
bag radius and a decrease of the parameters $Z_{B}$ for the heavier baryons.

Next we fit the quark-meson coupling constants $g_\sigma^q$, 
$g_\omega = 3g_\omega^q$ and $g_\rho = g_\rho^q$ for the nucleon to obtain 
the correct saturation properties of
the nuclear matter, $E_B \equiv   \epsilon/\rho - M = -15.7$~MeV at
$\rho~=\rho_0=~0.15$~fm$^{-3}$, $a_{sym}=32.5$ MeV, $K=257$ MeV and
$M^*=0.774 M$. We have 
$g_\sigma^q=5.957$, $g_{\omega N}=8.981$ and  $g_{\rho N}=8.651$.
We take the standard values for the meson masses, $m_\sigma=550$ MeV, 
$m_\omega=783$ MeV $m_\rho=770$ MeV. 

For the meson-hyperon coupling
constants we have opted for three sets discussed in the literature : 
set a) based on quark counting arguments we take 
$x_{\sigma B}=x_{\omega B}= x_{\rho B}=\sqrt{2/3}$ as in \cite{moszk};
set b) according  to \cite{gm91,Glen00} we choose the hyperon coupling 
constants constrained by the binding of the $\Lambda$ hyperon in nuclear 
matter, hypernuclear levels and neutron star masses
($x_\sigma=0.7$ and $x_\omega=x_\rho=0.783$) and assume that the couplings to 
the $\Sigma$ and $\Xi$ are equal to those of the $\Lambda$ hyperon;  
set c) based on the $SU(6)$ symmetry for the light quarks ($u,d$) counting 
rule \cite{pal} we take $x_{\sigma B}=x_{\omega B}=2/3$ and 
$x_{\rho \Lambda}=0, x_{\rho \Sigma}=2, x_{\rho \Xi}=1$.

\begin{figure}
\begin{center}
\includegraphics[width=9.cm]{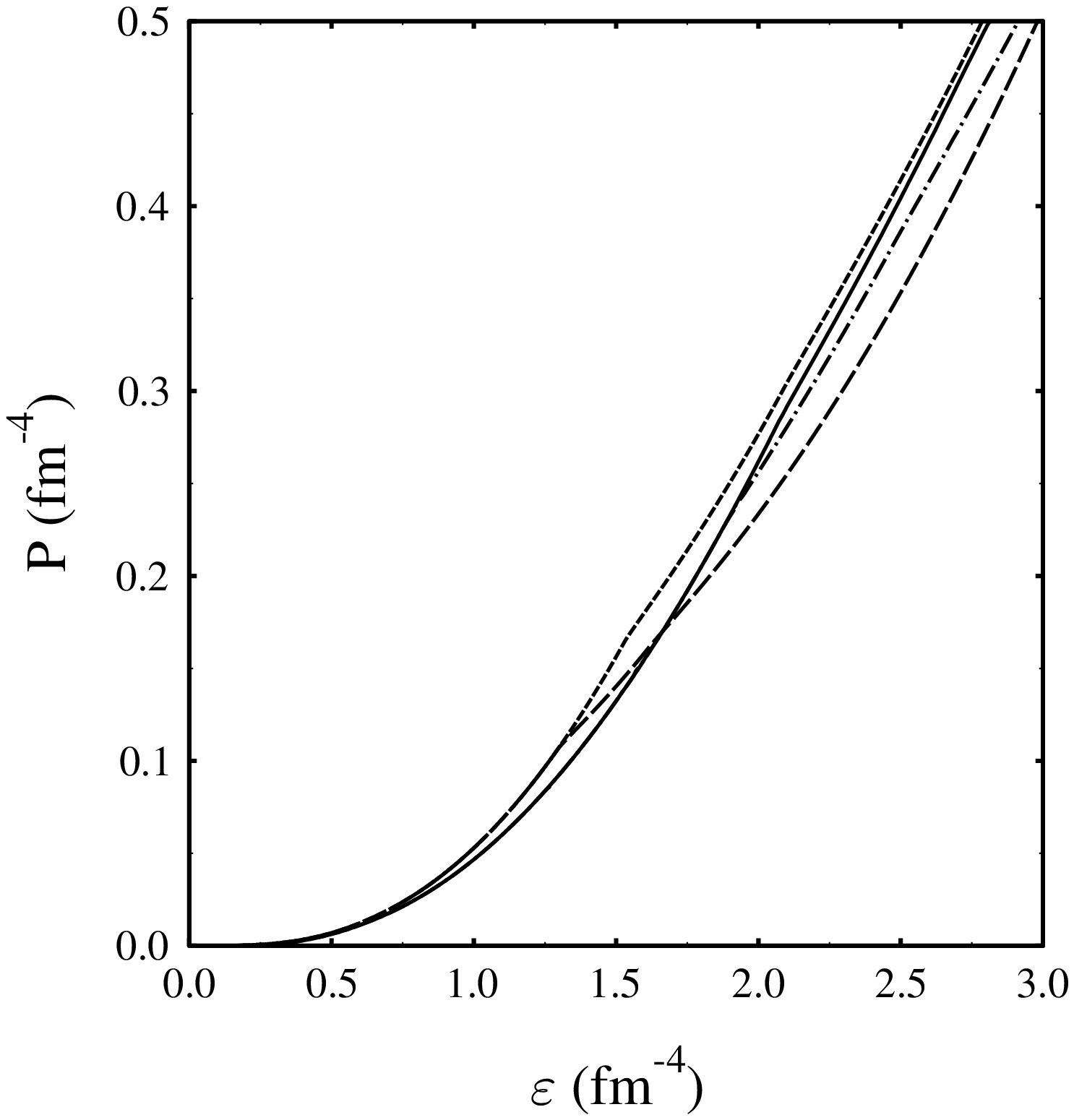} 
\end{center}
\caption{EOS obtained with QMC and set a) (solid line), QMC and set b) 
(dot-dashed line), NLWM and set a) (dashed line) and NLWM and set b) 
(dotted line) including hyperons.}
\label{nlwmVqmc}
\end{figure}

In Fig. \ref{nlwmVqmc} we have plotted the EOS obtained with the above 
parametrization of QMC and two choices of the hyperon couplings, 
set a) and b) together with the corresponding EOS obtained with NLWM with 
cubic and quartic sigma terms  ($K=$300 MeV and $M^*=0.7$ M). Although
the  NLWM-EOS is harder at low densities, it becomes softer at higher 
energies after the onset of the hyperons. 
This fact has
consequences on the behavior of the mixed phase and on the star properties.
Moreover, one can see that different choices of the meson-hyperon parameters
have a greater influence on the NLWM, for which the curves separate at lower
densities and become more distant one from the other at high densities, than 
in the QMC.

\section{Unpaired quark matter equation of state}

The possible existence of quark matter in the core of neutron stars is
an exciting possibility \cite{freedman}. Densities of these stars
are expected to be high enough to force the hadron constituents or
nucleons to overlap, thereby yielding quark matter. 
We take the quark matter equation
of state as in Ref. \cite{farhi,kapusta} in which $u$,$d$ and $s$
quark degrees of freedom are included in addition to electrons.
Up and down quark masses are set to 5.5 MeV and
the strange quark mass is taken to be 150 MeV. In chemical
equilibrium $\mu_d=\mu_s=\mu_u+\mu_e$. In terms of neutron and
electric charge chemical potentials $\mu_n$ and $\mu_e$, one has
\begin{equation}
\mu_u={1\over 3}\mu_n-{2\over 3}\mu_e,\quad
\mu_d={1\over 3}\mu_n+{1\over 3}\mu_e,\quad
\mu_s={1\over 3}\mu_n+{1\over 3}\mu_e.
\end{equation}

The pressure for quark flavor $f$, with $f=u,d$ or $s$ is \cite{kapusta}
\begin{eqnarray}
P_q&=&{1\over 4\pi^2} \sum_f \left[\mu_f k_f(\mu_f^2-2.5m_f^2)+
1.5m_f^4ln\left({\mu_f+k_f\over m_f}\right)\right],
\end{eqnarray}
where the Fermi momentum is $k_f=(\mu_f^2-m_f^2)^{1/2}$. 

For the leptons, the pressure reads
\begin{equation}
P_l= \frac{1}{3 \pi^2} \sum_l \int \frac{p^4 dp}
{\sqrt{p^2+m_l^2}}.
\label{pressl}
\end{equation}

The total pressure, including the bag constant B, which simulates confinement 
becomes
\begin{equation}
P= P_l+ P_q-B\ .
\end{equation}

There are only two independent chemical potentials $\mu_n$ and
$\mu_e$. $\mu_e$ is adjusted so that the matter is electrically
neutral, i.e. $\partial P/\partial \mu_e=0$.

\section{Color-flavor locked quark phase}
In this section we study the equation of state taking into consideration a 
CFL quark paired phase. We treat the quark matter as a Fermi sea of free quarks
with an additional contribution  to the pressure arising from the 
formation of the CFL condensates.

The CFL phase can be described with the help of the thermodynamical potential 
which reads \cite{ar}:
\begin{equation}
\Omega_{CFL}(\mu_q,\mu_e)=\Omega_{quarks}(\mu_q) + \Omega_{GB}(\mu_q,\mu_e) +
\Omega_{l}(\mu_e),
\end{equation}
where $\mu_q=\mu_n/3$ and 
\begin{equation}
\Omega_{quarks}(\mu_q)=\frac{6}{\pi^2} 
\int_0^{\nu} p^2 d p (p-\mu_q)
+\frac{3}{\pi^2} 
\int_0^{\nu} p^2 d p (\sqrt{p^2 + m_s^2}-\mu_q)
- \frac{3 \Delta^2 \mu_q^2}{\pi^2} +B,
\end{equation}
with  $m_u=m_d$ set to zero,
\begin{equation}
\nu=2 \mu_q - \sqrt{\mu_q^2 + \frac{m_s^2}{3}},
\end{equation}
$\Omega_{GB}(\mu_q,\mu_e)$ is a contribution from the Goldstone bosons arising
due to the chiral symmetry breaking in the CFL phase \cite{son,ar}:
 
\begin{equation}
\Omega_{GB}(\mu_q,\mu_e)=-\frac{1}{2} f_{\pi}^2 \mu_e^2 \left(1 - \frac{m_\pi^2}
{\mu_e^2} \right)^2,
\label{ogb}
\end{equation} 

where 
\begin{equation}
f_{\pi}^2=\frac{(21-8~~ ln 2)\mu_q^2}{36 \pi^2}
~~~,~~~
m_\pi^2= \frac{3 \Delta^2}{\pi^2 f_\pi^2} m_s (m_u + m_d),
\end{equation}
$\Omega_{l}(\mu_e)$ is the negative of expression (\ref{pressl}),
and the quark number densities are equal, i.e.,
\begin{equation}
\rho_u=\rho_d=\rho_s=\frac{\nu^3 + 2 \Delta^2 \mu_q}{\pi^2}.
\end{equation}
In the above expressions $\Delta$, the gap parameter, is taken to be 100 MeV 
\cite{ar}.

The electric charge density carried by the pion condensate is given by
\begin{equation}
Q_{CFL}=f_\pi^2 \mu_e \left(1 - \frac{m_\pi^4}{\mu_e^4} \right).\label{qcf}
\end{equation}
In the above thermodynamic potential, we have neglected the contribution
due to the kaon condensation which is an effect of  order  $m_s^4$  and thereby
small compared to the$\Delta^2\mu_q^2$ contribution to the thermodynamic 
potential for $\Delta\sim 100$ MeV.

\section{Mixed Phase and hybrid star properties}

We now consider the scenario of a mixed phase of hadronic and quark matter. 
In the mixed phase charge neutrality is imposed globally i.e. the quark and 
hadronic phases are not neutral separately
but rather, the system prefers to rearrange itself so that
\begin{equation}
\chi\rho_c^{QP}+(1-\chi)\rho_c^{HP}+\rho_c^{l}=0
\end{equation}
where $\rho_c^{QP}$ and $\rho_c^{HP}$ are the charge density of quark and 
hadron phase, $\chi$ is the volume fraction occupied by the quark phase,
$(1-\chi)$ is the volume fraction occupied by the hadron phase and $\rho_c^{l}$
is the lepton charge density. As usual, the phase boundary of the
coexistence region between the hadron and quark phase is determined
by the Gibbs criteria. The critical pressure and critical neutron and electron 
chemical potentials are determined by the conditions,

$$\mu_{HP,i}=\mu_{QP,i}=\mu_i, \,\, i=n,e, \quad T_{HP}=T_{QP},\quad P_{HP}(\mu_{HP},T)=P_{QP}
(\mu_{QP},T),$$
reflecting the needs of chemical, thermal and mechanical equilibrium, 
respectively.
The energy density and the total baryon density in the mixed phase read:
\begin{equation}
\varepsilon=\chi\varepsilon^{QP}+(1-\chi)\varepsilon^{HP}+\varepsilon^{l},
\end{equation}
\begin{equation}
\rho=\chi\rho^{QP}+(1-\chi)\rho^{HP}.
\end{equation}

Notice that in all equations above the quark phase (QP) can be either the UQM
or the CFL phase. The EOS for the mixed phase are then constructed. Once they 
are obtained, the properties of the neutron stars can be computed.
The equations for the structure of a relativistic  spherical and
static star composed of a perfect fluid were derived from Einstein's
equations by Oppenheimer and Volkoff \cite{tov}. They are
\begin{equation}
\frac{dP}{dr}=-\frac{G}{r}\frac{\left[\varepsilon+P\right ]\left[M+
4\pi r^3 P\right ]}{(r-2 GM)},
\label{tov1}
\end{equation}
\begin{equation}
\frac{dM}{dr}= 4\pi r^2 \varepsilon ,
\label{tov2}
\end{equation}
with $G$ as the gravitational constant and $M(r)$ as the enclosed gravitational
mass. We have used $c=1$.
Given an EOS, these equations can be integrated from the origin as an initial
value problem for a given choice of the central energy density, 
$(\varepsilon_0)$.
The value of $r~(=R)$, where the pressure vanishes defines the
surface of the star. We solve the above equations to study the structural 
properties of the star, using the EOS derived above. 

\section{Results and discussion}

In all figures shown, set a) for the meson-hyperon coupling constants was 
used, unless stated otherwise. We have omitted all results for the parameter 
sets b) and c) because they are very similar to the ones obtained with set 
a). Actually, all differences appear only in the hadron phase at densities 
where the mixed phase is already the dominant one. 
It is worth emphasizing that, as stated in sections III and IV, the $u$ and 
$d$ quark masses are different in the UQM and CFL models. 

\begin{table}
\begin{ruledtabular}
\caption{Bag constants for the baryons at the free space value $B_B^{1/4}$.
The third and forth columns are obtained for $B_B^{1/4}=210.854$ and the mass
of the quarks taken as $m_u=m_d=5.5$ MeV and $m_s=150$ MeV. 
The last two columns are for $B_B^{1/4}=211.303$ and the mass
of the quarks taken as $m_u=m_d=0$ MeV and $m_s=150$ MeV}
\label{tab1}
\begin{tabular}{cccccc}
Baryons& $M_B$ & $Z_B$ & $R_B$ & $Z_B$ & $R_B$ \\ \hline
N &939.0  & 4.00506 & 0.6 &3.98699 & 0.6\\
$\Lambda$ &1115.6  & 3.69005 & 0.62525 & 3.68029 & 0.62428\\
$\Sigma^+$ &1189.3  & 3.45577 & 0.63977 & 3.44628 & 0.63870 \\
$\Sigma^0$ &1192.5  & 3.40386 & 0.64038 & 3.43600 & 0.63931 \\
$\Sigma^-$ &1197.4  & 3.42970 & 0.64038 & 3.42024 & 0.64024 \\
$\Xi^0$ &1314.9  & 3.29260 & 0.65336 & 3.29188 & 0.65182 \\
$\Xi^-$ &1321.3  & 3.27173 & 0.65455 & 3.27105 & 0.65301 \\
\end{tabular}
\end{ruledtabular}
\end{table}
\begin{table}
\begin{ruledtabular}
\caption{Mixed star properties}
\label{tab2}
\begin{tabular}{ccccccc}
model & $B^{1/4}$ (MeV)& $M_{max}/M_\odot$ & 
$\varepsilon_0$ (fm$^{-4}$) & $\varepsilon_{min}$ (fm$^{-4}$) & 
$\varepsilon_{max}$ (fm$^{-4}$) \\ 
\hline
QMC+UQM & 180     &1.41 & 8.53 & 1.26 & 5.24 \\
QMC+UQM & 190     &1.58 & 5.52 & 1.63 & 7.02 \\ 
QMC+UQM & 200     &1.73 & 4.85 & 2.05 & 8.74 \\
QMC+UQM & 210.854 &1.85 & 4.68 & 2.73 & 10.57 \\
\hline
QMC+CFL & 190     & 1.32 & 12.56& 1.35 & 4.56 \\
QMC+CFL & 200     & 1.49 & 3.31 & 1.92 & 6.25 \\
QMC+CFL & 211.303 & 1.76 & 3.94 & 2.66 & 8.28 \\
\hline
NLWM+UQM & 180    & 1.40 & 7.38 & 1.17 & 4.62\\
NLWM+UQM & 190    & 1.64 & 4.58 & 1.81 & 6.06 \\
\end{tabular}
\end{ruledtabular}
\end{table}

\begin{figure}
\begin{center}
\begin{tabular}{cc}
\includegraphics[width=9.cm]{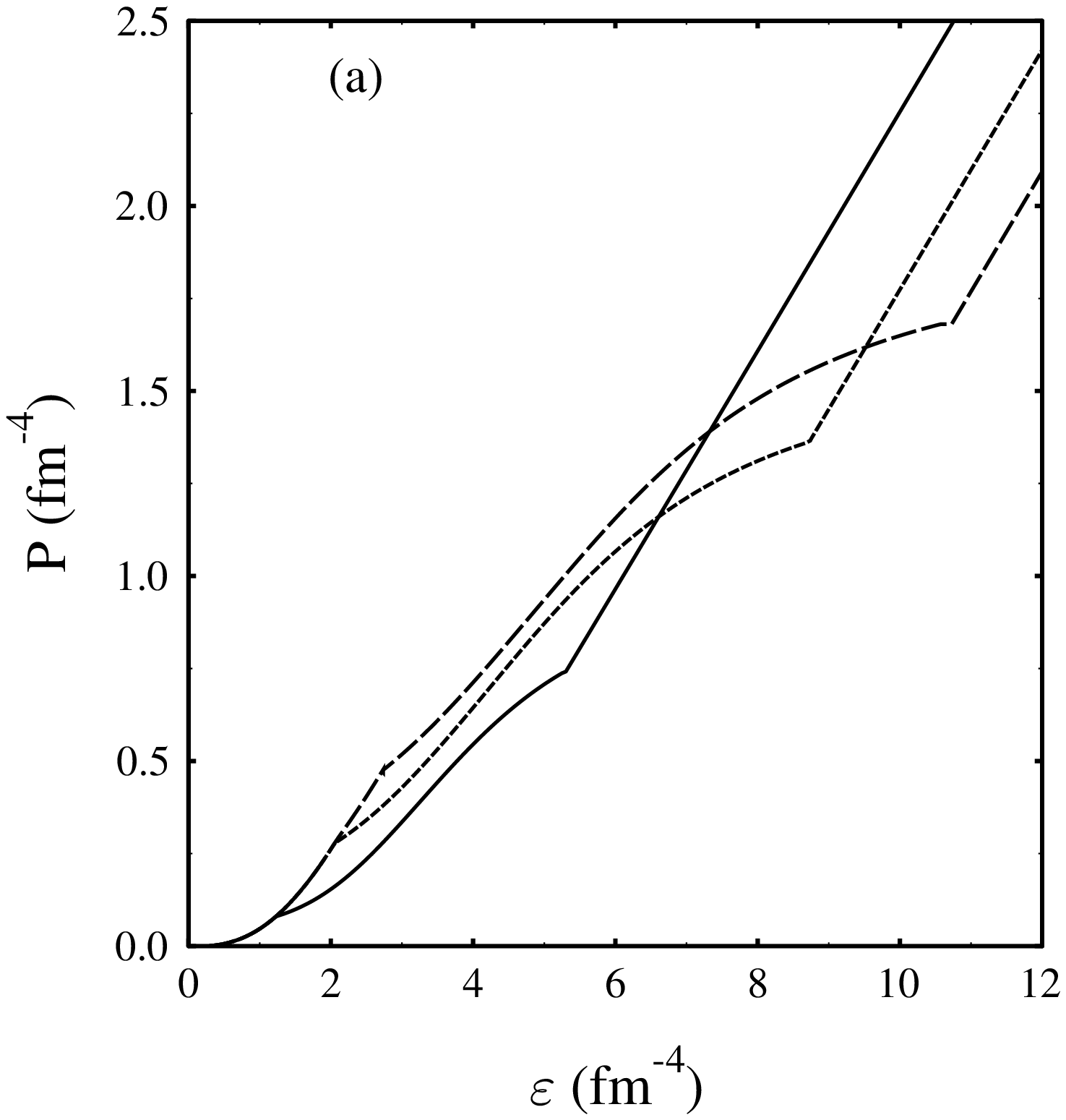} &
\includegraphics[width=9.cm]{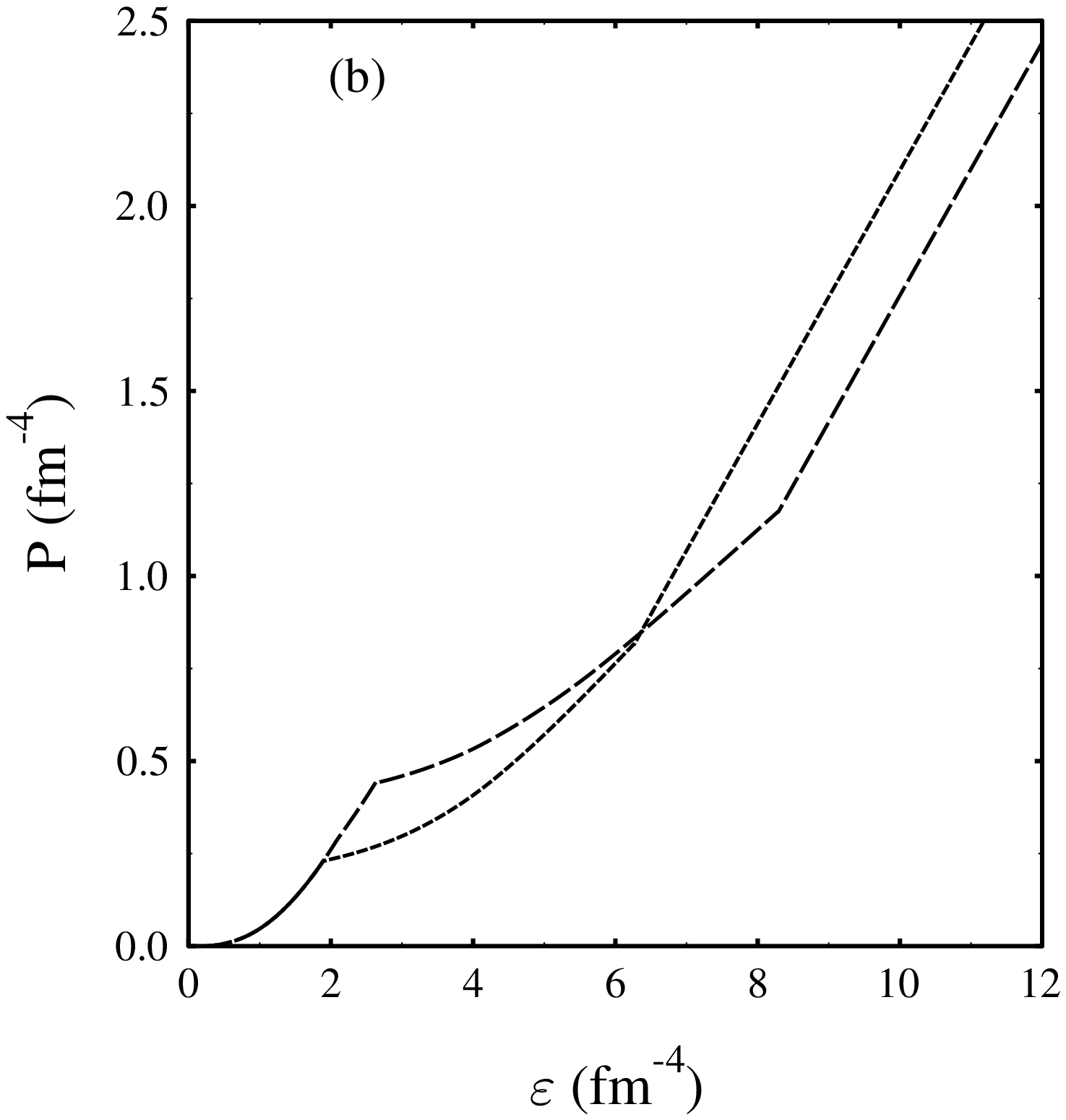}
\end{tabular}
\end{center}
\caption{EOS obtained with QMC plus UQM (a) for B$^{1/4}$=180 (solid line),
200 (short-dashed line) and 210.85 MeV (long-dashed line) and QMC plus CFL (b)
for B$^{1/4}$=200 (short-dashed line) and 211 MeV (long-dashed line) }
\label{eos}
\end{figure}

\begin{figure}
\begin{center}
\includegraphics[width=9.cm]{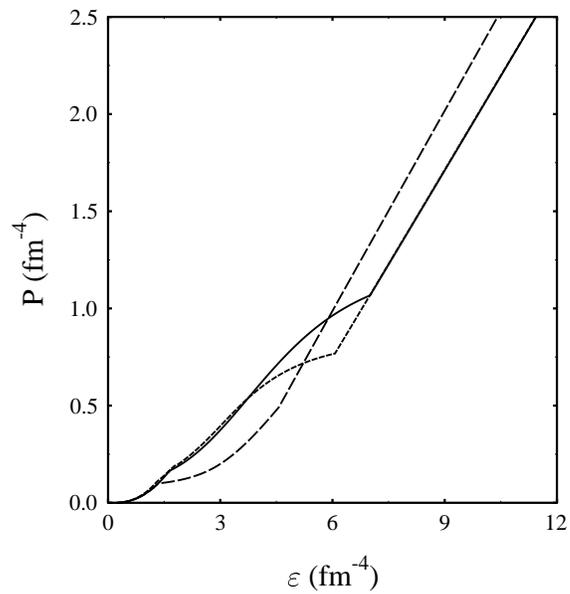}
\end{center}
\caption{EOS obtained with QMC plus UQM (solid line), 
QMC plus CFL (dashed line) and NLWM plus UQM (dotted line) with B$^{1/4}$=190 
MeV. Only here we have considered
$m_u=m_d=0$ MeV in both QMC models to compare the equations of state.}
\label{eos2}
\end{figure}

In figures \ref{eos}a) and \ref{eos}b) the EOS obtained with the unpaired
quark model and the color flavor locked phase are displayed for
different values of the Bag pressure $B$. The onset of the mixed phase and 
quark pure phase occurs at lower denities for smaller values of $B$. 
This effect has already been discussed in \cite{prak97} for the UQM 
description. A smaller value of $B$ gives a softer EOS in the mixed
phase because the onset of the mixed phase occurs at lower densities. However at higher 
densities, after the onset of the quark phase it becomes harder. The mixed 
phase shrinks with the decrease of the $B$ parameter in both quark models. 
This fact can be also observed in table \ref{tab2}, where the beginning and 
ending energy densities of the mixed phase are displayed in the last two 
columns. For the CFL, no mixed phase was found for B$^{1/4}$=180 MeV with $m_s=150$ MeV and $\Delta=100$ MeV, giving rise to a pure quark matter star. 
For the sake of comparison we have plotted the EOS 
for the same bag pressure for the UQM and CFL phase in figure \ref{eos2}. 
We include in the same figure  the
EOS obtained with NLWM plus UQM (short-dashed line).
One can see from this figure and table \ref{tab2} that
the mixed phase in the CFL appears and ends at lower energy densities than 
in the UQM. This result has the same qualitatively behavior as the ones shown 
in \cite{ar}. Comparing QMC and NLWM with UQM one can see that they behave
similarly at low densities. For $\epsilon>4.5 \mbox{fm}^{-4}$ the NLWM-EOS 
becomes softer in the mixed phase. This is due to the
fact that at higher densities the NLWM with hyperons becomes softer than the 
QMC-EOS, Fig. \ref{nlwmVqmc}.

\begin{figure}
\hspace*{-1cm}
\begin{tabular}{cc}
\includegraphics[width=9.5cm]{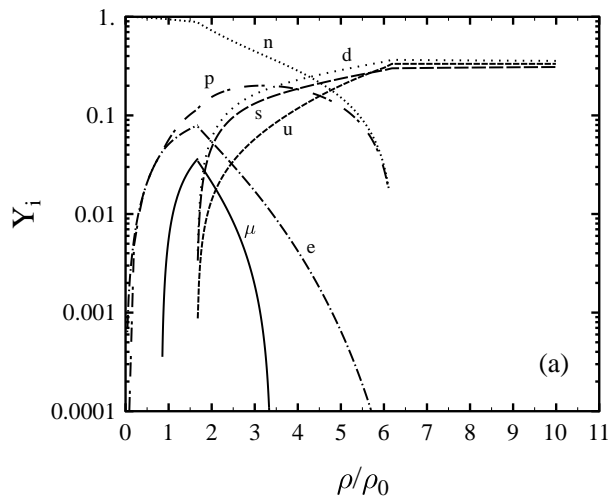}&
\includegraphics[width=9.5cm]{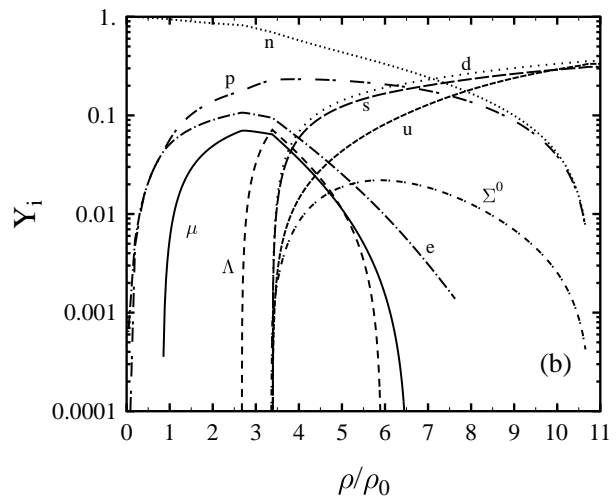} \\
(a)&(b)\\
\includegraphics[width=9.5cm]{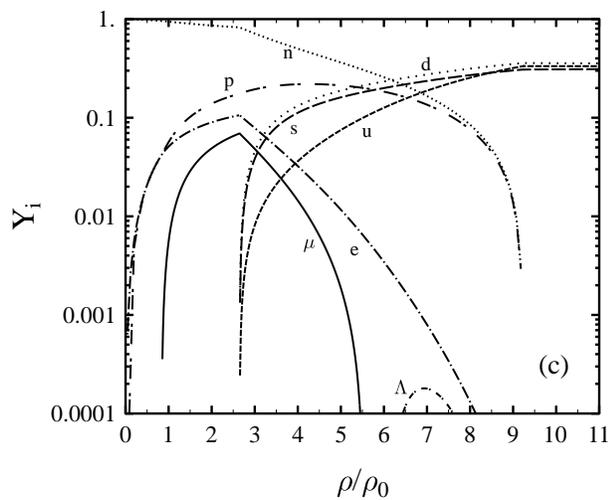}&
\includegraphics[width=9.5cm]{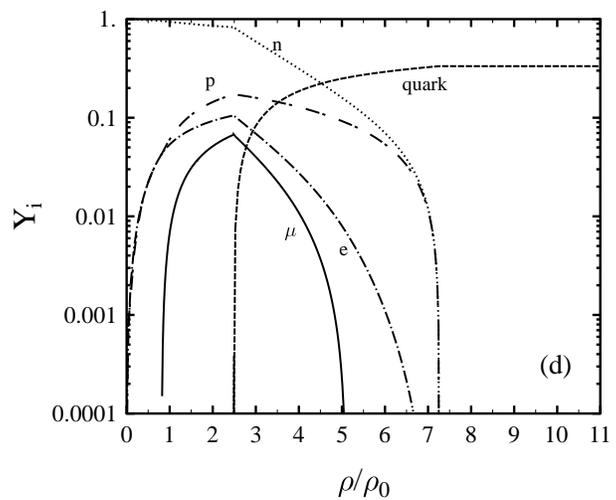}\\
(c)&(d)
\end{tabular}
\caption{Particle fractions, $Y_i=\rho_i/\rho$, for $i=$ baryons, leptons and 
quarks, obtained with the QMC+ 
UQM (a) for B$^{1/4}$=180 MeV, QMC+UQM (b) for B$^{1/4}$=210.85 MeV  
QMC+UQM (c) for B$^{1/4}$=200 MeV, and QMC+CFL (d) for B$^{1/4}$=200 MeV}
\label{yi}
\end{figure}

In figure \ref{yi} the particle population for the baryons, leptons and quarks 
are shown for the UQM and different bag pressures and for the CFL with one
chosen Bag value. Hyperons only appear if the  bag constant is very high, 
in our case $B^{1/4}=210.85$ MeV. This value  corresponds the $B_B$ 
value of the bag used in the QMC model for the hadronic phase.
A different behavior was obtained 
with NLWM in \cite{recente1} where the hyperons are present both for 
$B^{1/4}=180$ and 190 MeV.
As already discussed, both the  mixed phases and the quark phase appear at
lower densities for lower $B$ values.
For the CFL phase and $B^{1/4}$ =200 MeV, the quarks appear 
at $2.67$ nuclear saturation density. In \cite{ar}, for a bag pressure of
$B^{1/4}$ =190 MeV, the quarks appear at 2.149 $\rho_0$. We can see that
our results are compatible with the ones shown there, obtained within a NLWM 
formalism with $K=240$ MeV and $M^*=0.78M$. 
While in figures \ref{yi}c), $u,d$ and $s$ quark populations come out
different, in
figure \ref{yi}d), because of the imposition of equal quark densities in the 
CFL  model, they are forced to be equal. Also, the $e^-$ population, which 
disappears at the onset of the quark phase in all the EOS studied, in the 
CFL model, goes to zero at a much lower density than in the UQM model, because 
the onset of a pure quark phase occurs at lower densities.

\begin{figure}
\begin{center}
\includegraphics[width=9.cm]{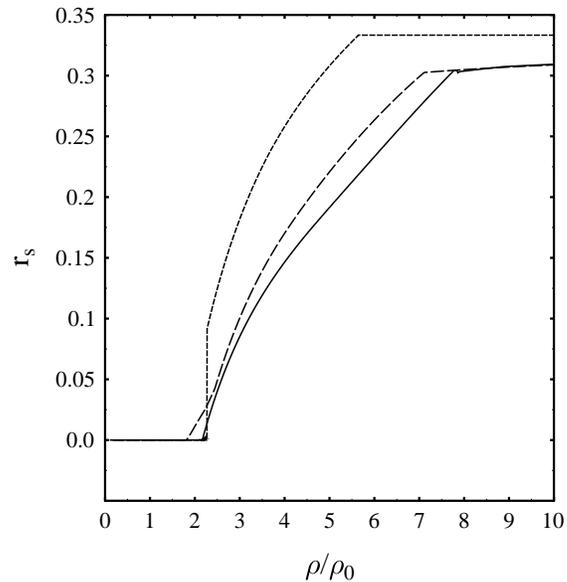}
\end{center}
\caption{Strangeness content obtained with QMC plus UQM (solid line),
QMC plus CFL (dotted line) and NLWM plus UQM (dashed line) for 
B$^{1/4}$=190 MeV.}
\label{rs}
\end{figure}

As already discussed in \cite{recente1,recente2}, the presence of strangeness 
in the core and crust of neutron and proto-neutron stars has important 
consequences in  understanding some of their properties. 
In figure \ref{rs} we show the strangeness fraction  defined as
\begin{equation}
r_s=\chi \, r_s^{QP}+ (1-\chi)\, r_s^{HP}
\end{equation}
with
$$r^{QP}_s=\frac{\rho_s}{3\rho}, \quad r^{HP}_s=
\frac{\sum_B |q^B_s|\rho_B}{3\rho}, $$
where $q_s^B$ is the strange charge of baryon $B$, for the different models 
discussed in this work. In all cases the 
strangeness fraction rises steadily. If the UQM is used,
at the onset of the pure quark phase 
it has reached  30 per cent of the baryonic matter. Although the amount of 
strangeness
varies in the mixed phase, it is the same in the pure quark phase independently
of the model used to describe the hadron phase. Nevertheless, if the CFL model
is used for the quark phase, as a result of the equal quark densities 
imposition, the strangeness content reaches 1/3 of the total baryonic matter.
Comparing QMC and NLWM with UQM,  we conclude the strangeness content 
increases faster when the NLWM is used. This is due to the fact that in this 
model the hyperons also contribute.

In table \ref{tab2} we show the values obtained for the maximum mass of a 
neutron star as function of the central density for some of the EOS studied 
in this work with UQM and CFL. Different bag parameters are used. We can see
that the maximum mass of the star increases and its central energy decreases
with increasing $B$.  This result agrees with the fact that a larger $B$ 
value corresponds to a harder EOS at high densities as shown in Fig. \ref{eos}.
For $B^{1/4}=180$ MeV (QMC+UQM) and $B^{1/4}=190$ MeV (QMC+CFL), the 
central density of the star lies outside of the range of the mixed phase, 
an indication of a star with a quark core. In the last case we get a very 
high central density due to the very soft EOS this parametrization gives rise 
to. In fact the predicted maximum mass for a hybrid star within this 
parametrization is too low as discussed below.
In all the  other cases discussed the central density of the star is 
always within the mixed phase. We have also added some results obtained with 
the non-linear Walecka model \cite{recente1} instead of the QMC for the 
hadron phase for the sake of comparison. For the same $B$, the maximum mass 
is about the same, 
the central core is also made up of quarks and the mixed phase starts and 
finishes at lower densities. In \cite{ar}, 
the authors have found maximum masses around 1.6 $M_\odot$ for maximum $B$ 
values of $B^{1/4}=185$ MeV. For these $B$ values, our results 
come out at the same order.
 
The radius of the maximum mass star is sensitive to the low density EOS. In 
order to calculate the radius and to plot it versus the star mass, we
have used the results of Baym, Pethick and Sutherland \cite{bps} for low 
baryonic densities. 

\begin{figure}
\begin{center}
\begin{tabular}{cc}
\includegraphics[width=9.cm]{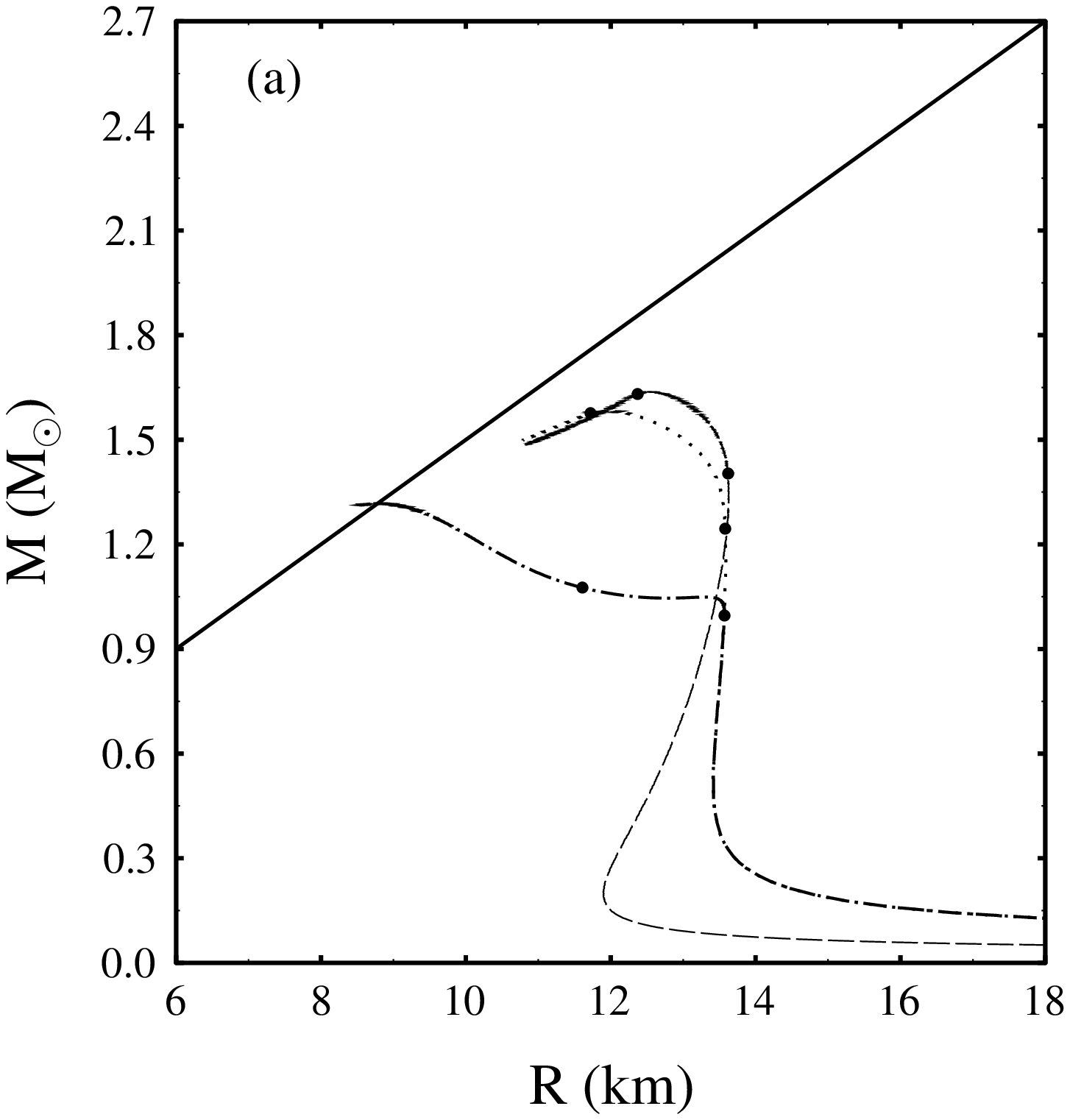} &
\includegraphics[width=9.cm]{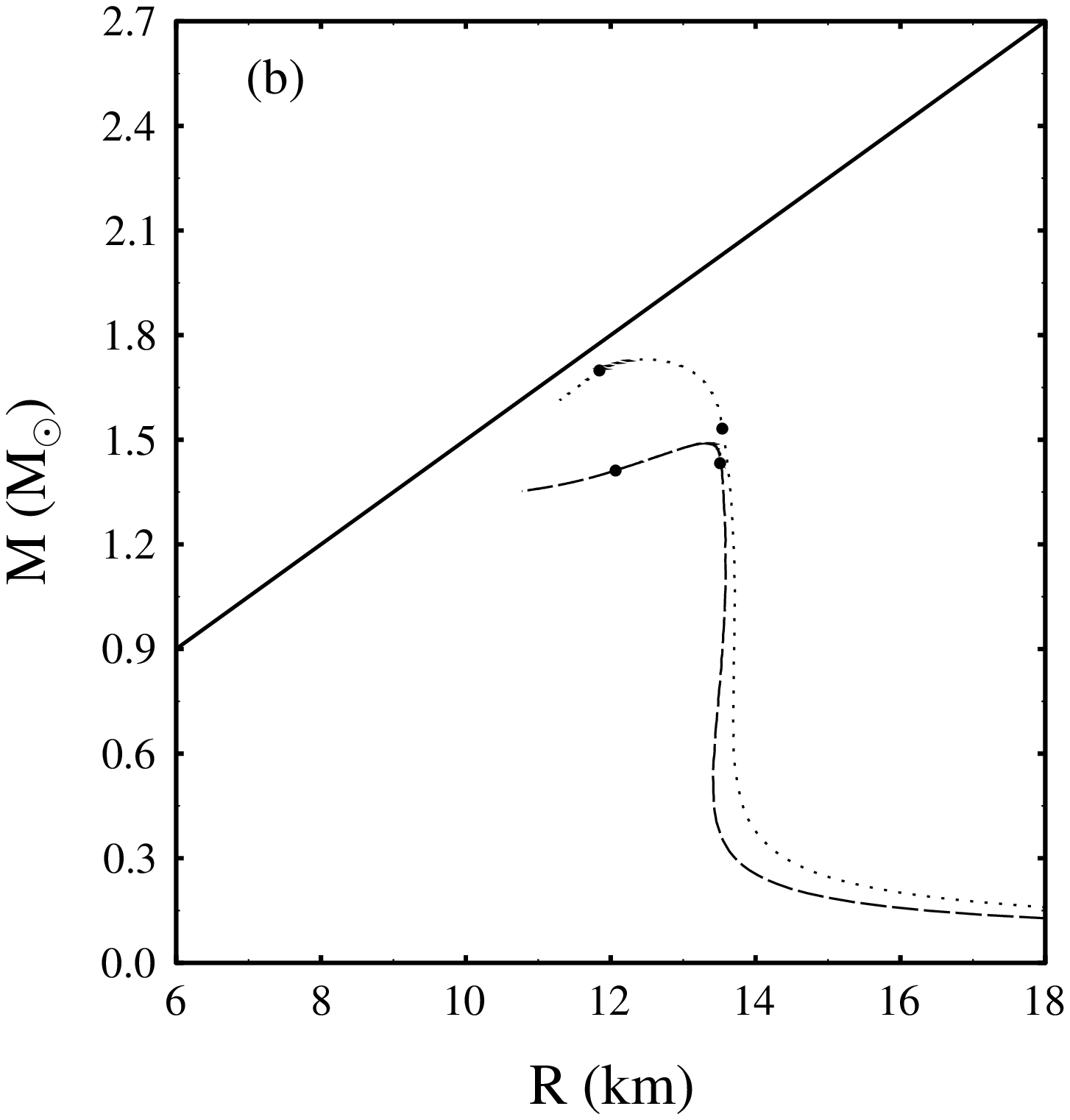}\\
(a)&(b) 
\end{tabular}
\end{center}
\caption{Neutron star mass versus radius obtained with the
(a) QMC plus UQM (dotted line), 
QMC plus CFL (dash-dotted line) and NLWM plus UQM (long-dashed line)  for 
B$^{1/4}$=190 MeV, 
(b) QMC plus UQM (dotted line) and QMC plus CFL (dashed line) for 
B$^{1/4}$=200 MeV. The dots indicate the beginning and end of the mixed 
phases.}
\label{tov}
\end{figure}

In Fig.  \ref{tov} the mass of the family of stars obtained with QMC is 
plotted  in terms of  their radii for both quark models used and 
$B^{1/4}=190$ and 200 MeV. We also include the
family of stars obtained with NLWM plus UQM with $B^{1/4}=190$ MeV. 
The radius of the stars within QMC is 11.95 Km ($B^{1/4}=190$ MeV) and  
12.47 Km  ($B^{1/4}=200$ MeV) for UQM and 8.79 Km ($B^{1/4}=190$ MeV) and 
13.31 Km ($B^{1/4}=200$ MeV) for CFL.  For the NLWM family of stars we get 
$R=12.53$ Km  ($B^{1/4}=190$ MeV).  
Some conclusions can be drawn. Comparing QMC and NLWM for $B^{1/4}=190$ MeV, 
the maximum mass of a stable star is similar, for both models. The fact that  
NLWM- $M_\odot$ is larger shows that the main
contribution to the star comes from the less dense regions, where the 
NLWM-EOS is harder than the QMC-EOS (see Fig. \ref{eos2}).  We consider now 
the families of stars obtained within QMC for both $B$ values and using UQM 
and CFL. The quark contribution becomes more important for the smaller $B$ 
values mainly for the CFL results. In particular,  for $B^{1/4}=190$ MeV we 
get a quite small maximum $M_\odot$ and a corresponding small radius. A 
maximum mass of 1.32 $M_\odot$ is too small for accounting for the presently 
known radio-pulsar masses \cite{tc}, even after corrections due to rotation.  
Similar numbers where obtained with the NLWM plus CFL in \cite{ar} for 
slightly different values of $B$ and $m_s$. In fact, since these
maximum mass stars have a small hadron exterior region the properties of the 
star is mainly determined by the quark model used.  
It is also clear from Fig. \ref{tov} that the star properties are sensitive 
to the $B$ value, in particular within the CFL formalism. 
Taking  $B^{1/4}=200$ MeV the contribution of the mixed phase to stable stars 
is much smaller, restricted to the core of the star. In Fig. \ref{tov}b) 
the properties of the stars which only contain hadronic matter do not 
coincide because we have taken $m_u=m_d=5.5$ MeV for the QMC parametrization 
for the results QMC plus UQM and zero otherwise.

A straightforward method of determining neutron stars
properties is by measuring the gravitational
redshift of spectral lines produced in neutron star photosphere which
provides a direct constraint on the mass-to-radius ratio.
Recently a redshift of 0.35 from
three different transitions of the spectra of the X-ray binary EXO0748-676
was obtained in \cite{cottam}. 
This redshift corresponds to $M_\odot/R(km)=0.15$. 
In Fig. 6 we have added the line corresponding to this constraint and only 
the EOS for QMC plus CFL  with  
B$^{1/4}$=190 MeV barely satisfies this constraint. 
In fact, the above constraint excludes all the EOS with hyperons, quarks or 
obtained within a relativistic mean-field approach, as compiled in 
\cite{lat01}.

\section{Conclusions}

In the present paper we have studied the EOS for neutron stars using 
both the unpaired quark matter based on the MIT Bag model and the CFL  phase 
for  describing the quark phase and a relativistic mean-field 
quark-meson coupling description in which quarks interact via the exchange of 
$\sigma$-, $\omega$- and $\rho$- mesons for the hadron phase.

We have compared the results obtained within the QMC model with the ones 
obtained within the NLWM. For similar properties at nuclear saturation we 
conclude that contrary to NLWM, with the QMC the hyperons only appear for 
a very high value of the bag constant. Also hyperons make the NLWM-EOS much 
softer than the QMC-EOS.  For the hyperon meson couplings we have used 
three choices and verified that they did not have any effect on the onset of 
hyperons. The none appearance of hyperons affects the
variation of the strangeness content of the star with density: except for 
the quark phase we get higher fractions with the NLWM.

For the bag pressure parameter $B$ we have  used three different values, 
which produce different EOS and consequently the properties of the stars
are dependent on them. Small $B$ values give stable stars with a quark
core. However, for a given  gap constant $\Delta$ we obtain a phase 
transition to a deconfined CFL phase only if $B$ is greater than a critical 
value.  For lower values we get a EOS of pure quark
matter, and pure quark matter stars. 
For $\Delta=100$ MeV we should have $B^{1/4} \ge 185$ MeV. 
For $B^{1/4} =190$ MeV and $m_s=150$ MeV, with QMC plus CFL we are not able 
to obtain stable stars with masses equal to most of radio-pulsar masses known. 
We have also shown that just one of the EOS we have studied 
satisfies the constraint imposed by the recently measured  
redshift of 0.35 from three different transitions of the spectra of the X-ray binary EXO0748-676\cite{cottam}. A more
systematic study has to be done in order to verify whether a set of acceptable
parameters for the QMC+ MIT
bag model gives a family of stars which contains the measured value.

In \cite{pal}, strange meson fields,
namely the scalar meson field $f_0(975)$ and the vector meson field
$\phi(1020)$, were also considered in order to reproduce the observed 
strongly attractive $\Lambda \Lambda$ interaction. They have shown that the 
introduction of these strange mesons makes the EOS harder due to the repulsive 
effect of  the $\phi(1020)$, meson.  A harder EOS for the hadronic matter 
gives rise to a onset of the mixed phase at
lower densities and a smaller mixed phase. The inclusion of these 
mesons and their influence on the properties of the stars are under 
investigation.

It has been shown that the effect of temperature on the maximum mass of 
stable stars is small compared to the effect of neutrino trapping 
\cite{prak97,recente1,recente2}. Therefore, it would be interesting to 
include neutrino trapping even at $T=0$ MeV
and impose  leptonic number conservation in the models used here and check
the properties of the arising stars.

\section*{ACKNOWLEDGMENTS}
This work was partially supported by CNPq (Brazil), CAPES (visiting researcher
project), CAPES(Brazil)/GRICES (Portugal) under project 003/100 and 
FEDER/FCT (Portugal) under the project POCTI/35308/FIS/2000.

\newpage
\end{document}